\documentclass[12pt]{iopart}

\usepackage{amssymb}
\usepackage[utf8]{inputenc}
\usepackage[english]{babel}
\usepackage{graphicx}
\usepackage{dcolumn}
\usepackage{bm}
\usepackage{url}
\usepackage[normalem]{ulem}
\usepackage{color}

\newcommand{\sign}{\mathop{\rm sign}\nolimits}
\newcommand{\sinc}{\mathop{\rm sinc}\nolimits}

\begin{document}

\title{One Dimensional $^1$H, $^2$H and $^3$H}
\author{A. J. Vidal$^1$, G. E. Astrakharchik$^1$, L. Vranje\v{s} Marki{\'c}$^2$, and J. Boronat$^1$}
\address{
$^1$ Departament de F\'\i sica, Campus Nord B4-B5, Universitat Polit\`ecnica de Catalunya,
E-08034 Barcelona, Spain\\
$^2$ Faculty of Science, University of Split, HR-21000 Split, Croatia
}
\date{\today}

\begin{abstract}
The ground-state properties of one-dimensional electron-spin-polarized hydrogen $^1$H, deuterium $^2$H, and tritium $^3$H are obtained by means of quantum Monte Carlo methods.
The equations of state of the three isotopes are calculated for a wide range of linear densities.
The pair correlation function and the static structure factor are obtained and interpreted within the framework of the Luttinger liquid theory.
We report the density dependence of the Luttinger parameter and use it to identify different physical regimes:
Bogoliubov Bose gas, super-Tonks-Girardeau gas, and quasi-crystal regimes for bosons;
repulsive, attractive Fermi gas, and quasi-crystal regimes for fermions.
We find that the tritium isotope is the one with the richest behaviour.
Our results show unambiguously the relevant role of the isotope mass in the properties of this quantum system.
\end{abstract}

\maketitle


\section{\label{sec:1}Introduction}

The quest for  observing Bose-Einstein condensation in cold gases was accomplished in Nobel-prize winning experiments~\cite{BEC95Cornell,BEC95Ketterle} in 1995. Since then, alkali gases have proven to be an extremely versatile experimental tool, as the interaction strength can be tuned by Feshbach resonance while the use of
optical lattices permits to create highly controllable geometries.
Reduced dimensionality might lead to highly non-trivial phenomena in quantum systems.
It was shown~\cite{Girardeau60} by Marvin Girardeau in 1960 that in one dimension the wave function of bosons with strong repulsion can be
mapped to a wave function of non-interacting fermions.
In this system, known as Tonks-Girardeau (TG) gas, bosons acquire many fermionic properties revealing the intricate relation between quantum statistics in one dimension.
Later on, it was proposed~\cite{Astrakharchik05} that a gas with even stronger correlations than in a TG gas (super Tonks-Girardeau (sTG) gas) can be obtained by crossing rapidly the confinement induced resonance~\cite{Olshanii98}.
Tonks-Girardeau gas was successfully realized and observed with
$^{87}$Rb~\cite{Kinoshita04,Paredes04} atoms and TG and sTG gases with
$^{135}$Cs~\cite{Haller09,Haller11} atoms.

Before the breakthrough progress was achieved with alkali atoms, the most studied candidate for observing Bose-Einstein condensation in cold gases was electron-spin-polarized hydrogen.
A group led by Thomas Greytak and Daniel Kleppner at MIT began their quest for the quantum degeneracy of atomic hydrogen already in 1978.
A magnetic trap with evaporative cooling was one of the techniques used by the group which later was adopted by the alkali gas laboratories.
It took two decades to finally reach~\cite{Kleppner98} the Bose-Einstein condensation in atomic hydrogen, in 1998.
We might hope that the experimental techniques developed since then for alkali gases can be backported to hydrogen for creating clean one-
systems of hydrogen and its heavier isotopes, deuterium and tritium. 
In particular, the possibility for formation of a tritium condensate using its broad Feshbach resonance was suggested in Ref.~\cite{blume}.

Our additional motivation is that, out of all atoms, hydrogen has the simplest structure and it is a very basic fundamental question to find out its properties in reduced dimensionalities.
A further advantage of considering polarized hydrogen is that its interatomic potential is exactly known from works by Kolos, Jamieson, Dalgarno, and Wolniewicz~\cite{kolos1968,POT2,potboro}.
As the model potential is the only input of quantum Monte Carlo methods, we can benefit of its power for producing very accurate quantitative
results~\cite{HDT3,bevslic2009quantum,xu2002structural,bevslic2013quantum}.

The main goal of the present study is to find the ground-state properties of hydrogen $^1$H, deuterium $^2$H, and tritium $^3$H isotopes of spin-polarized hydrogen in one dimension.
In Sec.~\ref{sec:2}, we introduce the quantum Monte Carlo method used for the simulation.
Section~\ref{sec:3} comprises the main results of the work including the energy of the systems, their structure properties and the determination of the Luttinger parameter as a function of the density for each of the three isotopes.
In Sec.~\ref{Sec:Discussion} we make a comparison with other quantum one-dimensional systems.
Finally, the main conclusions are drawn in Sec.~\ref{Sec:Conclusions}.

\section{\label{sec:2}Method}

We use the diffusion Monte Carlo method\cite{Guardiola,boronat1994monte} to study one-dimensional hydrogen $^1$H, deuterium $^2$H, and tritium $^3$H at zero temperature.
The Hamiltonian of the system is
\begin{equation}
\hat{\mathcal{H}}=-\frac{\hbar^2}{2m_{H}}\sum_{i=1}^N\Delta_i+\sum_{i<j}^N V\left(\left|x_i-x_j\right|\right) \;,
\label{Eq:H}
\end{equation}
where $x_i,\ i=\overline{1,N}$ denote the positions of the $N$ atoms.
We consider exactly the same interactions for all isotopes so that the distinguishing factor is the mass $m_{1H}=1.00794u$ (hydrogen), $m_{2H}=2.01410u$ (deuterium), and $m_{3H}=3.01605u$ (tritium).
The triplet $b~^3\Sigma_u^+$ interaction potential $V(x)$ is obtained by interpolating data from highly accurate 'ab-initio' calculations by Jamieson {\it et al.} (JDW)~\cite{POT2} and by smoothly connecting them to an attractive $r^{-6}$ tail, term that comes from the interaction of induced electric dipoles at large distances~\cite{yan}.
The JDW potential exhibits an attractive well with a minimum of $-6.49$~K at $x=4.14$~\AA\, and has a repulsive hard wall at short distances.
This potential was used to predict ground-state properties and stability of bulk and clusters of hydrogen, deuterium, and tritium\cite{Beslic08,Beslic09,Stipanovic11,Beslic13}.
We study the 1D homogeneous system by applying periodic boundary conditions to a box of length $L$.
We are interested in the microscopic properties of the system in the thermodynamic limit at a given linear density $\rho=N/L$.
In the following, we stick to conventional units of Angstrom for the distance and Kelvin for the energy.

The DMC method solves in a stochastic way the imaginary-time Schr\"odinger equation of a many-particle system.
To reduce the variance of the statistical estimations the method works with importance sampling.
This technique, which is widely used in any Monte Carlo calculation, relies in the present case on the introduction of a guiding wave function to drive the sampling to regions where one reasonably knows that the statistical weight is higher.
In our study, we use a Jastrow correlation factor $\Psi_B$ for bosons and $\Psi_F$ for fermions, given by
\begin{eqnarray}
\Psi_B(x_1,\dots,x_N)&=&\prod_{i<j}^Nf_2(|x_i-x_j|) \;,  \nonumber \\
\Psi_F(x_1,\dots,x_N)&=&\prod_{i<j}^Nf_2(|x_i-x_j|)\sign\left(x_i-x_j\right)
\;. \label{Eq:wf}
\end{eqnarray}
In Eq.~(\ref{Eq:wf}), $f_2(x)$ is chosen as the two-body scattering solution at short distances, $x<R_{\textrm{par}}$, and follows the phononic asymptotic law $|\sin(\pi x/L)|^{1/K_{\textrm{par}}}$ at large distances~\cite{Reatto67,Haldane81},
$x>R_{\textrm{par}}$.
We note that the guiding wave function~(\ref{Eq:wf}) becomes exact in the limit of low densities, when both the short-range and long-range parts become equal to $f_2(x) = |\sin(\pi x/L)|$.
This limit describes a Tonks-Girardeau gas for the bosonic $^1$H and $^3$H isotopes and an ideal Fermi gas for the fermionic one, $^2$H, since when the interparticle distance is large enough the contribution of the potential energy vanishes and then the energy is fully kinetic.
There are two variational parameters in the guiding wave function~(\ref{Eq:wf}), namely $R_{\textrm{par}}$ and
$K_{\textrm{par}}$.
The matching distance $R_{\textrm{par}}$ is optimized by minimizing the energy in a variational Monte Carlo (VMC) calculation.
The physical meaning of $K_{\textrm{par}}$ is that of the Luttinger parameter, which we chose consistently with the equation of state.
All the calculations are performed with $N=20$ particles which proved to be enough for reaching a reasonable description of the thermodynamic limit in one dimension.
Finally, we worked with pure estimators~\cite{pures} for the calculation of the static structure factor and pair distribution function in order to eliminate any residual bias coming from the guiding wave function.

\section{\label{sec:3}Results}

The energy and diagonal properties of the three hydrogen isotopes are independent of the statistics and/or polarization of the atoms due to Girardeau's mapping~\cite{Girardeau60}.
The mapping relies on having both a one-dimensional system and hard-core interactions, conditions that 1D hydrogen,
deuterium, and tritium satisfy.
According to Girardeau's mapping, there is a simple relation between fermionic $\Psi_F$ and bosonic $\Psi_B$ wave
functions, namely $\Psi_B=|\Psi_F|$.
As the interatomic potential is the same for the three isotopes, the mass is the crucial factor that controls their different behaviour.

\begin{figure}
\begin{center}
\includegraphics[width=0.7\textwidth]{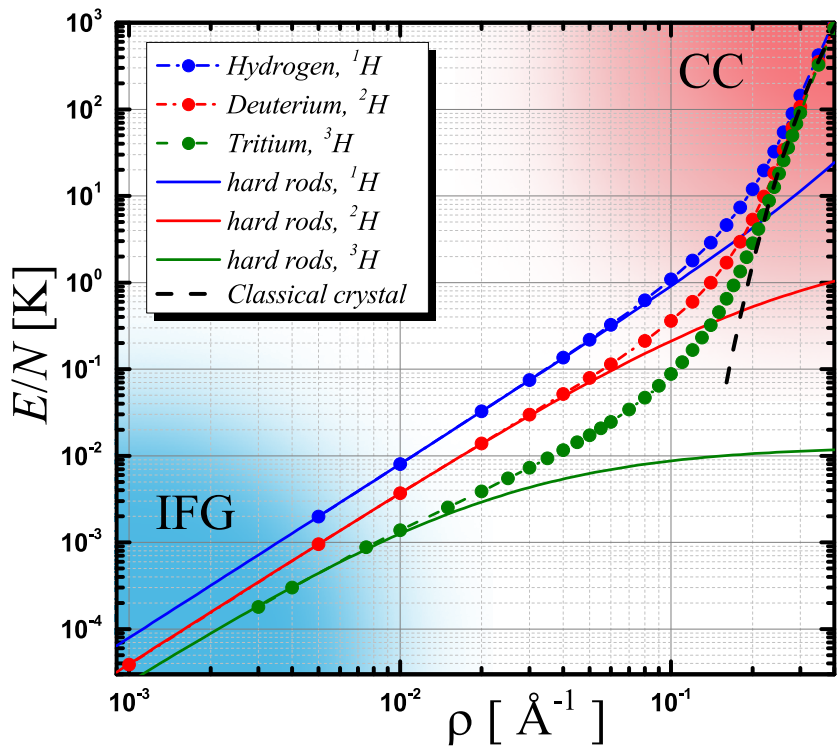}
\caption{Log-log plot of the equation of state (EoS) for the three isotopes of hydrogen:
hydrogen (upper curve), deuterium (middle curve), and tritium (lower curve).
Circles, DMC energy;
dashed lines, guide to an eye;
solid lines in the dilute regime,
energy of hard-rod gas, Eq.~(\ref{Eq:E:HR}) with corresponding $s$-wave scattering length and isotope mass;
dashed line at high densities, energy of the classical crystal, $E_{IC}/N$ Eq.~(\ref{Eq:HC}).
We indicate the ideal Fermi gas (IFG) and classical crystal (CC) areas by shaded areas.}
\label{Fig:EoS}
\end{center}
\end{figure}

\begin{figure}
	\begin{center}
		\includegraphics[width=0.7\textwidth]{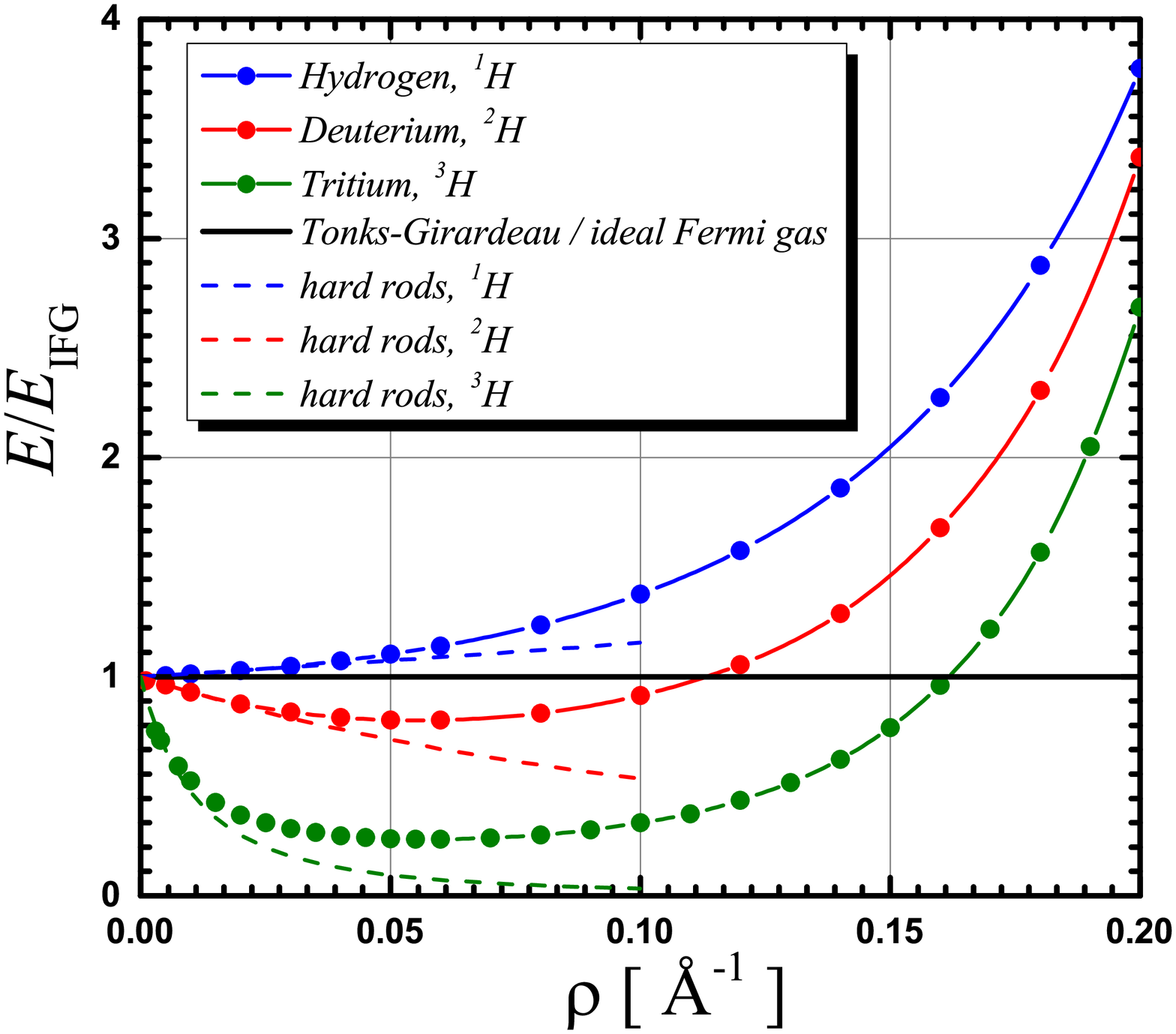}		
		\caption{Comparison of the energy per particle for the three hydrogen
			isotopes normalized by the energy of the corresponding ideal Fermi gas.
			Symbols, DMC energy (hydrogen, deuterium, tritium from top to bottom);
			solid lines connecting symbols, guide to an eye.
			In the limit $\rho\to 0$ we recover the energy of an ideal Fermi/Tonks-Girardeau gas, shown by a solid horizontal line $E = E_\mathrm{IFG}$, where $E_\mathrm{IFG}$ is given by Eq.~(\ref{Eq:E:IFG}) with the corresponding isotope mass.
			Dashed lines in the dilute regime show the energy of hard-rod gas, Eq.~(\ref{Eq:E:HR}) with corresponding $s$-wave scattering length and isotope mass.
		}
		\label{fig:E:vs:IFG}
	\end{center}
\end{figure}

In Fig.~\ref{Fig:EoS}, we report the density dependence of the ground-state energy for all three isotopes.
The energy per particle is a monotonously increasing function of the density.
No local minima in the energy are found, so none of the isotopes is able to form a self-bound liquid phase.
It is worth noticing that tritium, having the largest mass, exhibits a tendency to form an inflection, as one can see in Fig.~\ref{Fig:EoS}.
In fact, in three dimensions bulk tritium at zero temperature is a liquid with an equilibrium density $\rho_0=0.0075$~\AA$^{-3}$~\cite{tritiumleandra}.
Nevertheless, there are distinct physical regimes which can be identified from the equation of state and the distribution functions.
In the dilute limit, $\rho\to 0$, the energetic and diagonal properties are that of an ideal Fermi gas (IFG) for all three isotopes.
The energy of the IFG has a quadratic dependence on the density,
\begin{equation}
\frac{E}{N} = \frac{\pi^2\hbar^2\rho^2}{6m}\;,
\label{Eq:E:IFG}
\end{equation}
and appears as a straight line on the double logarithmic plot of Fig.~\ref{Fig:EoS}.
In this density range, it is the same as the energy of the TG gas.
For intermediate densities, around $\rho\sim 0.05$~\AA$^{-1}$, deuterium and tritium behave like Bose gases, tritium being
the one with the more marked behaviour, see Fig.~\ref{fig:E:vs:IFG}.
Around this density, the attractive long-range part of the interaction contributes significantly to the potential energy.
As a result, the total energies of $^2$H and $^3$H are below the energy of the ideal Fermi gas.
Instead, the energy of hydrogen $^1$H is always larger than the ideal Fermi gas one.

\begin{figure}
\begin{center}
\includegraphics[width=0.7\textwidth,angle=0]{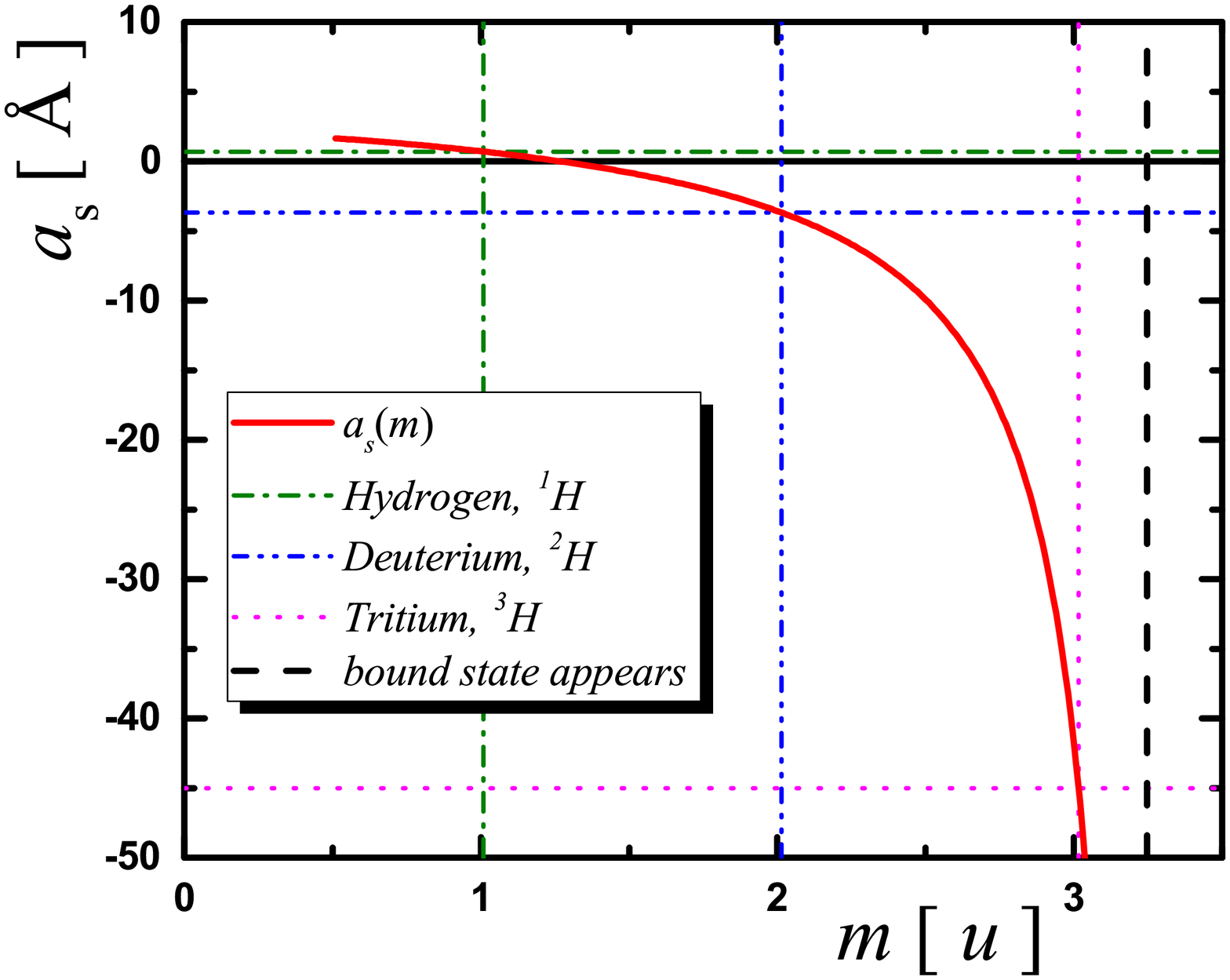}
\caption{Mass dependence of the $s$-wave scattering length $a_s$.
Solid line, solution of the two-body scattering problem;
dash-dotted lines, $^1$H with $m = 1.00794u$ and $a_s = 0.70$\AA;
dash-dot-dotted lines, $^2$H with $m = 2.0141u$ and $a_s = -3.69$\AA;
dotted lines, $^3$H with $m = 3.01605u$ and $a_s = -45.0$\AA;
long-dashed line, critical value $m_{c} = 3.25u$ at which a two-body bound state appears.}
\label{Fig:as(m)}
\end{center}
\end{figure}

The deviations from the Tonks-Girardeau/ ideal Fermi-gas regime can be analyzed using scattering theory.
The rapid long-range decay of the interatomic potential permits to describe it at low densities with the scattering phase shift or the $s$-wave scattering length $a_s$.
In Fig.~\ref{Fig:as(m)} we report the dependence of $a_s$ on the mass of the atom.
For hydrogen the $s$-wave scattering length is positive, $a_{1H} = 0.70$\AA, making the interaction effectively similarly to that of hard rods (HR) or super-Tonks-Girardeau branch in contact-interacting gases.
At the same time, the small value of $a_s^{1H}$ means that the deviations from the Tonks-Girardeau energy will be relatively weak.
For deuterium the $s$-wave scattering length changes sign and is equal to $a_s^{2H} =-3.69$\AA.
The negative value makes the interaction potential be similar to that of the contact interaction, $V(x)=g\delta(x)$, with the usual 1D relation between the coupling constant $g$ and the $s$-wave scattering length $g = -2\hbar^2/(ma_s) > 0$.
The large value of the $s$-wave scattering length suggests that the deviations from the ideal Fermi gas will happen at much smaller values of the density.
For the tritium the density dependence will become very prominent as the $s$-wave scattering is very large, $a_s^{3H}=-45.0$\AA. Indeed, the tritium mass is very close to the threshold value $m = 3.27 u$ at which a bound state enters.

The equation of state of a hard-rod gas is obtained from the energy of the ideal Fermi gas, Eq.~(\ref{Eq:E:IFG}) by subtracting the excluded volume of the hard rods $Na_s>0$ from the system size, $L\to L - N a_s$, resulting in\cite{Girardeau60}
\begin{equation}
\frac{E}{N}
= \frac{\pi^2\hbar^2\rho^2}{6m(1-\rho a_s)^2}
= \frac{\pi^2\hbar^2\rho^2}{6m^2} [1+2\rho a_s+3(\rho a_s)^2 + O((\rho a_s)^3)] \;,
\label{Eq:E:HR}
\end{equation}
Interestingly, Eq.~(\ref{Eq:E:HR}) also describes the energy of a dilute gas with $a_s<0$, i.e. for the contact interaction Lieb-Liniger gas.
Its equation of state can be obtained from Bethe {\it ansatz} approach and coincides with Eq.~(\ref{Eq:E:HR}) even if the $s$-wave scattering length has an opposite sign, compared to hard rods with $a_s>0$, making ``excluded volume'' correction effectively ``increase'' the available phase space, $L\to L + N |a_s|$.
Physically such a coincidence reflects the continuity of the Lieb-Liniger and super-Tonks-Girardeau branches with the differences appearing only in $O((\rho a_s)^3)$ terms in Eq.~(\ref{Eq:E:HR}) where the effective range of the potential enters which is different for the contact- and hard-rod potentials\cite{PhysRevA.81.013612}.

By comparing the DMC results with Eq.~(\ref{Eq:E:HR}) we find a perfect agreement.
As expected from Eq.~(\ref{Eq:E:HR}), the departure from the ideal Fermi gas behavior, corresponding to straight lines in Fig.~\ref{Fig:EoS} happens at smaller density for tritium and at the largest density for hydrogen.
The corrections are best observed by analyzing the ratio of the energy and the energy of the ideal Fermi gas, $E/E_{IFG}$, shown in Fig.~\ref{fig:E:vs:IFG}.
The deviations from the IFG value are positive for hydrogen (hard-rod or super-Tonks-Girardeau type) and negative for deuterium and tritium (Lieb-Liniger type).

For even larger densities, the repulsive part of the potential makes the system become more rigid and leads to a rapid increase in the energy.
At $\rho\sim 0.11$~\AA$^{-1}$ for deuterium and $\rho\sim 0.16$~\AA$^{-1}$ for tritium, the energy per particle becomes larger than that of the IFG and, if the density is increased even more, the energy diverges quickly.
These effects can be conveniently seen in Fig.~\ref{fig:E:vs:IFG} where we plot the energy in terms of the energy of
the ideal Fermi gas.

For high densities, for example $\rho=0.3$~\AA$^{-1}$, the total energy is dominated by the potential energy of the hard-core repulsion and the harmonic crystal theory can be applied.
We note that it is an unusual feature of one-dimensional physics that while strictly speaking the system always remains in a gas phase, its properties still might be correctly described by a crystal.
The quasi-crystal description is applicable to one-dimensional gases interacting with dipolar\cite{Arkhipov05,Astrakharchik08,Astrakharchik09},
$1/r^2$\cite{Astrakharchik06} and Coulomb\cite{Astrakharchik11} potentials.
The harmonic crystal energy $E_{HC}$ is computed as the sum of the potential energy of a perfect crystal $E_{IC}$ and the zero-point motion
\begin{equation}
\frac{E_{HC}}{N}=\frac{E_{IC}}{N}+\frac{1}{\ell_{BZ}}\int_{BZ}\frac{\hbar\omega(k)}{2}\;dk \; ,
\label{Eq:HC}
\end{equation}
where $\ell_{BZ}$ is the size of the first Brillouin zone. The excitation spectrum is obtained from the classical Newton equations of motion for each atom (see Appendix),
\begin{equation}
\omega(k)=\sqrt{\frac{4}{m}\sum_{n=1}^{\infty} \left[\frac{\partial^2V}{\partial x^2}
\bigg|_{\frac{n}{\rho}} \sin^2\left(\frac{k n}{2\rho}\right)\right]} \ .
\end{equation}

We find that the high-density regime can be described by the harmonic crystal approach.
It is interesting to note that at comparable densities, one-dimensional helium cannot be yet approximated by the HC theory\cite{Astrakharchik14}.
In fact, HC approach is applicable when the repulsive hard-core part of the interaction potential provides a much larger contribution than the attractive tails.
The dominant contribution comes from the potential energy of a perfect classical lattice and depends only on the interaction potential $V(x)$ and is thus mass independent.
This is the reason why the equations of state for all isotopes approach each other for high densities.
By keeping only the exponential repulsive core in the interaction potential, $V(x) = V_0 \exp(-\varkappa |x|)$, we find that the energy at high densities diverges as
\begin{equation}
\frac{E^{(0)}}{N} = \frac{V_0}{e^{\varkappa/\rho}-1}
\label{Eq:HC0:exp}
\end{equation}
There are two types of correction to Eq.~(\ref{Eq:HC0:exp}).
The first has a classical nature and comes from the attractive tail of the interaction potential.
This correction is negative and is mass independent.
The second correction has a quantum nature and corresponds to the energy of the zero-point motion.
This correction is positive and has a weak $1/\sqrt{m}$ dependence on the isotope mass, see Eq.~(\ref{Eq:HC1:exp}).

The structural properties change significantly across different physical regimes.
In the following, we analyze the behaviour of the two-body distribution function $g(r)$ and the static structure factor $S(k)$.

\begin{figure}
\begin{center}
\includegraphics[width=0.7\textwidth]{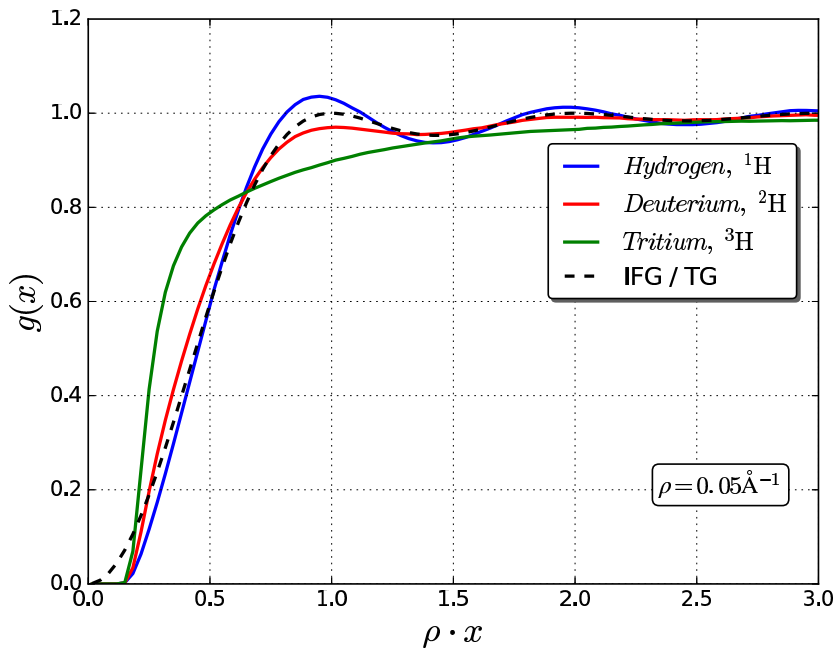}
\caption{Two-body distribution function $g(r)$ for the three isotopes at the intermediate density, $\rho=0.05$~\AA$^{-1}$.
Solid lines, DMC results for the hydrogen, deuterium and tritium (decreasing the height of the peak);
dashed line, ideal Fermi gas, Eq.~(\ref{Eq:g2:IFG}).}
\label{Fig:g2:rho}
\end{center}
\end{figure}

In the very dilute limit, the two-body distribution function, which is proportional to the probability to find two particles at distance $x$, approaches that of an ideal Fermi gas, given by
\begin{equation}
g_\mathrm{IFG}(x)=1-\sinc^2\left(k_F x\right)
\label{Eq:g2:IFG}
\end{equation}
where the Fermi wave number is $k_F=\pi\rho$.
For the fermionic $^2$H isotope it is quite natural that at low densities the Friedel-like oscillations are formed at the Fermi wave number. Still, even for the bosonic $^1$H and $^3$H isotopes the oscillations appear again at $k=k_F$, as the hard-core repulsion plays the role of an effective Pauli exclusion principle, resulting in $g(0) = 0$.
While the Fermi energy depends on the isotope mass, the Fermi wave number $k_F$ is entirely fixed by the density.
By rescaling distances in units of the linear density, the distribution functions of all three isotopes coincide for $\rho\to 0$
and reproduce $g_\mathrm{IFG}(x)$, Eq.~(\ref{Eq:g2:IFG}).
We report $g(r)$ for an intermediate density in Fig.~\ref{Fig:g2:rho}.
Friedel-like oscillations for $^1$H and $^2$H are visible as peaks centered at distances $\rho r=1,2,\dots$, physically corresponding to multiples of the average interparticle distance.
Out of all isotopes, bosonic $^1$H has the highest peak, which is above the IFG/TG value.
Such strong correlations are manifestations of the super-TG regime (a similar effect is also seen in the energy, Fig.~\ref{fig:E:vs:IFG}).
Fermionic deuterium $^2$H experiences Friedel-like oscillations, while the height of the peak is below the $\max_k S(k)=1$ value of the IFG gas, because at this density $^2$H starts to depart from the IFG model (see Fig.~\ref{fig:E:vs:IFG}) showing a behaviour typical to an attractive Fermi gas.
The most dramatic effect of the isotope mass is observed for bosonic $^3$H, where the Friedel-like oscillations are not visible at all at this density.
Instead, the attractive part of the interaction together with the large mass make tritium behave similarly to a weakly interacting Bose gas.
In fact, the shapeless structure of $g(r)$ is typical to Bogoliubov theory of a weakly interacting Bose gas.
It is interesting to note, that in tritium there is an enhanced probability to find two atoms at short distances.
This can be formally shown by recalling that the static structure factor is related by Fourier transform
$S(k) = 1-\int e^{ikr} (g(r)-1) dr$
to the $g(r)$.
Due to phononic low-lying excitations, $S(k=0)=0$, resulting in the condition that the area between the correlation function and the long-range asymptotic value must be preserved, $\int(g(r)-1)dr=1$.
From that it immediately follows that that the long-range suppression in $g(r)$ for tritium results in an enhanced probability to find two-particles at short distances compared to the other isotopes.

\begin{figure}
\begin{center}
\includegraphics[width=\textwidth]{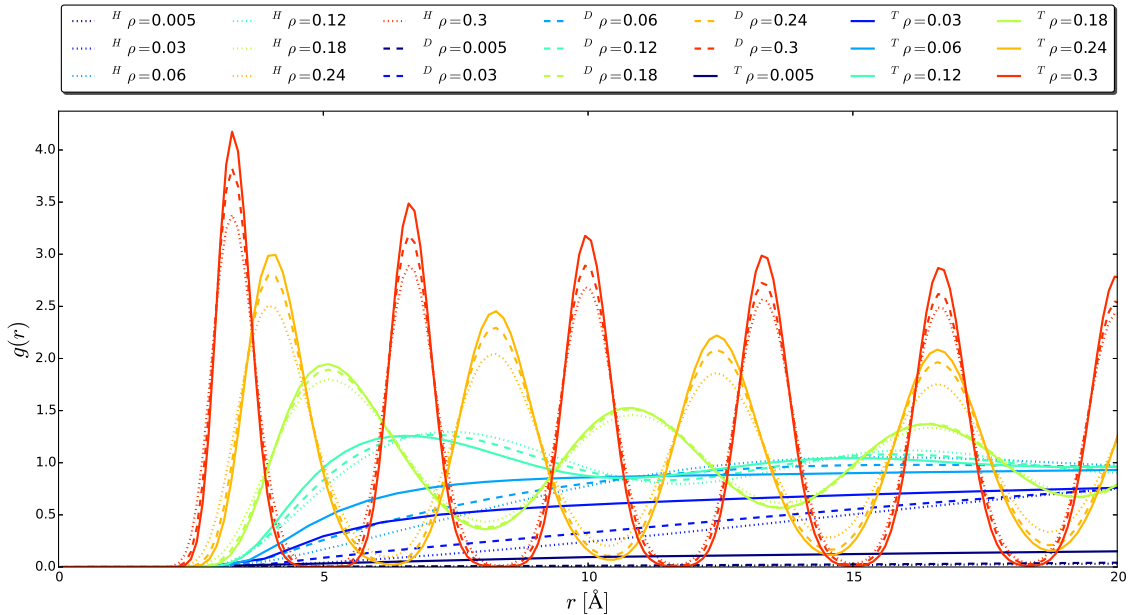}
\caption{Two-body distribution function of the three isotopes at densities $\rho$~\AA =0.3; 0.24; 0.18; 0.12; 0.06; 0.03; 0.005 (decreasing the height of the first peak).
The solid line, tritium T;
the dashed line, deuterium D;
the dotted line, hydrogen H. }
\label{Fig4}
\end{center}
\end{figure}

The short-range behavior is dominated by the hard-core repulsion.
In Fig.~\ref{Fig4} we show the two-body distribution functions on a linear scale.
The atoms cannot approach each other to distances smaller than few Angstroms.
This induces strong quantum correlations in the ground state.
The larger the density is, the larger is the potential energy.
In a certain sense the system becomes more classical.
The amplitude of the Friedel-like oscillations becomes larger and the atoms get more localized.
Note that the Friedel oscillations appear on the scale of the $1/k_F$ and are best seen when two-body distribution function is analyzed as a function of $\rho x$, like in Fig.~\ref{Fig:g2:rho}.
Instead, units of $\rho$~\AA\, are appropriate for studying the short-range behavior of $g(x)$ while for the smallest reported densities the Friedel oscillations appear at distances larger than reported in Fig.~\ref{Fig4}.
A quasi-crystal is eventually formed with the density profile of atoms at the equilibrium positions being almost Gaussian.
The lower mass of hydrogen results in a larger kinetic energy compared to the other isotopes and so the Gaussian distribution around the equilibrium positions is wider.
On the contrary, the peaks are the sharpest for tritium which has the heaviest mass.
The largest mass of tritium makes it effectively more classical at very high densities compared to other isotopes;
its equation of state is closest to the classical prediction, see Fig.~\ref{Fig:EoS}.

Quantum fluctuations destroy diagonal long-range order, even at zero temperature, and then no true crystal can be formed.
This can be deduced from the Luttinger liquid theory, which provides the following long-range expansion of the two-body distribution function\cite{Haldane81}
\begin{equation}
g(x)=1-\frac{K}{2\left(k_Fx\right)^2}+\sum_{\ell=1}^{\infty}A_{\ell}
\frac{\cos\left(2\ell k_Fx\right)}{\left|k_Fx\right|^{2\ell^2 K}}\;,
\label{Eq:g2:LL}
\end{equation}
where $K$ is the the Luttinger parameter and $A_{\ell}$ are coefficients.
The amplitude of oscillations decays as a power law at large distances and no true crystal is formed.
The Luttinger parameter $K$, which serves as an input characteristic for the phenomenological Luttinger liquid approach,
and the $A_{\ell}$ coefficients can be obtained only from a full many-body description which in our work is done by means of the diffusion Monte Carlo method.
The Luttinger parameter $K$ governs the long-range properties of the system.
Furthermore, its knowledge permits to exploit effective Hamiltonian theories and to extract important information on how the system behaves when an external field is applied.
For example, for $K<1$ the system gets pinned by a single impurity\cite{Glazman92} and no transmissibility is possible through a weak link\cite{Kane92,Furusaki93};
for $K<3/2$ the random disorder induces
localization~\cite{Giamarchi88,Ristivojevic12}; for $K<2$ a commensurate optical lattice induces transition to a Mott-insulator
phase~\cite{Bloch08}.
In the following, we extract the Luttinger parameter $K$ from the static structure factor $S(k)$.

\begin{figure}
\begin{center}
\begin{center}
\includegraphics[width=0.6\textwidth]{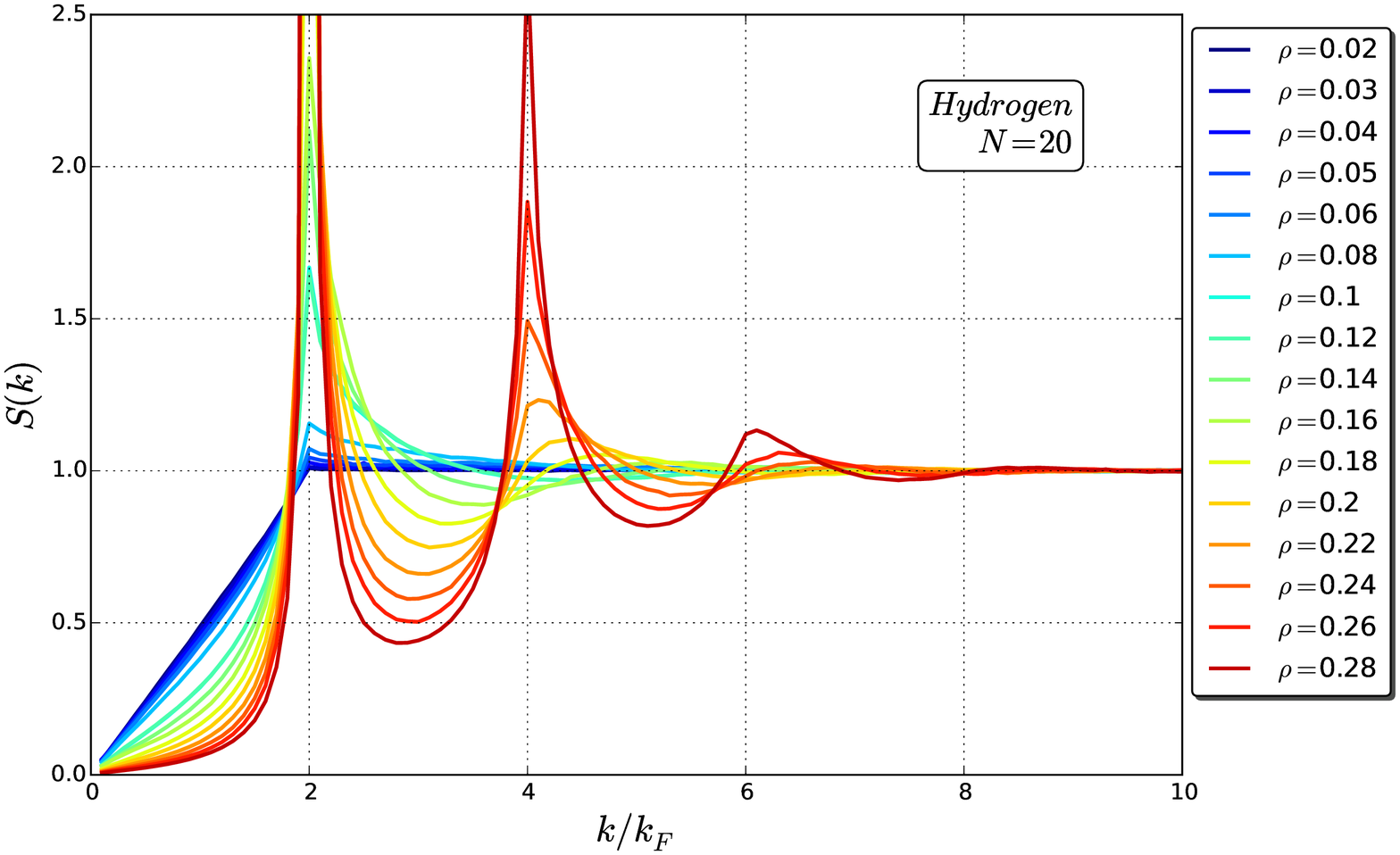}
\includegraphics[width=0.6\textwidth]{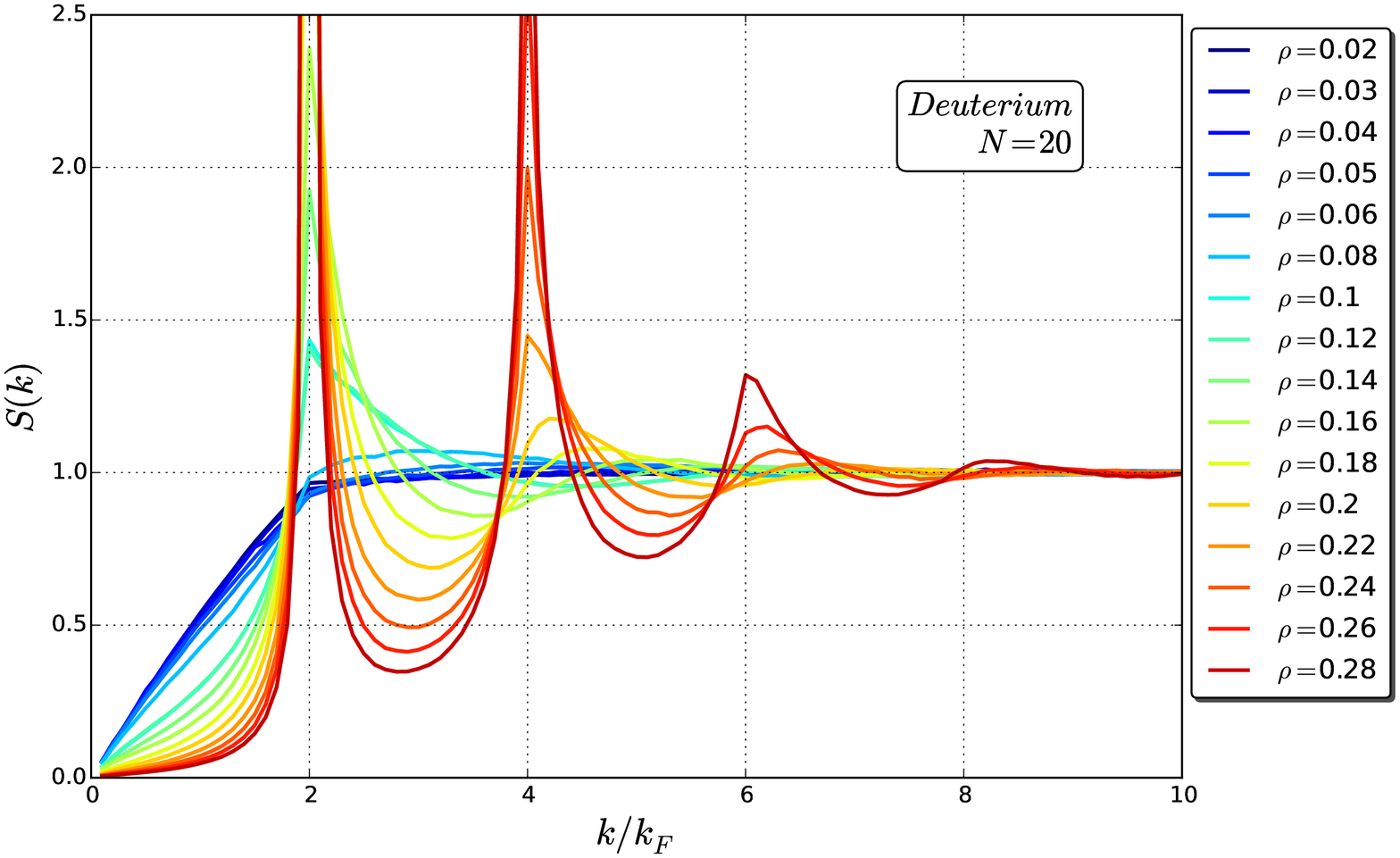}
\includegraphics[width=0.6\textwidth]{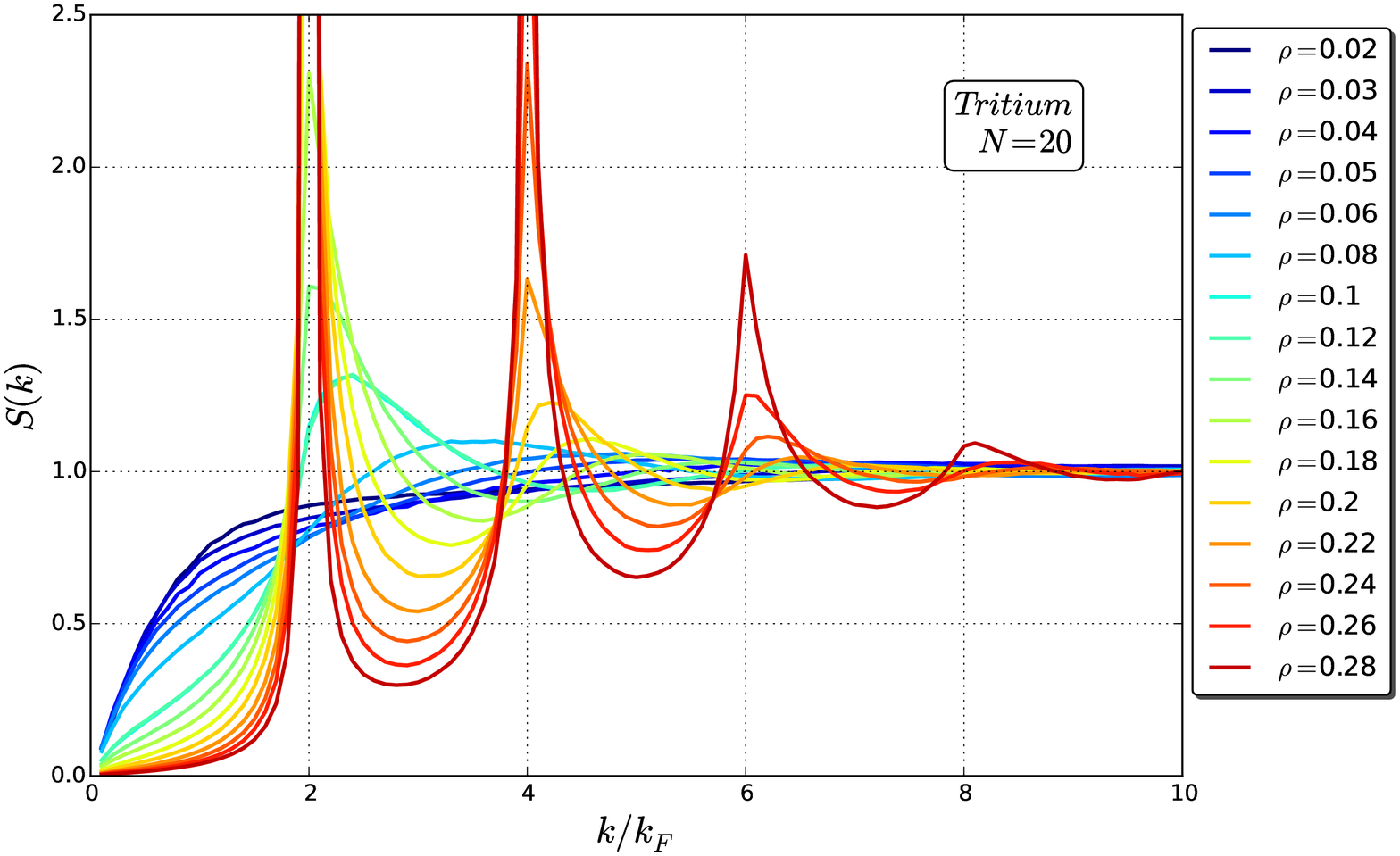}
\end{center}
\caption{Static structure factor comparison for the three isotopes at different densities.
Top (5a) is for hydrogen, middle (5b) for deuterium and bottom (5c) for tritium.
The different colours reflect the different densities.}
\label{Fig:Sk}
\end{center}
\end{figure}

The density dependence of the static structure factor $S(k)$ is shown in Fig.~\ref{Fig:Sk} for the three isotopes.
In the ultra-dilute regime it has the ideal-Fermi gas shape, with a linear low-momentum slope extending up to
$|k|=2k_F$ and followed by a constant $S(k)=1$ value for larger momenta.
At higher densities, the low-momentum behavior remains linear in $k$ and can be written as
\begin{equation}
S(k)=\frac{\hbar |k|}{2mc}\qquad k\to 0
\label{Eq:Sk:phonons}
\end{equation}
where $c$ is the speed of sound.
For the ideal Fermi gas the speed of sound is given by the Fermi velocity $c=v_F=\hbar \pi \rho / m$.

The linear behavior~(\ref{Eq:Sk:phonons}) at small $k$ reflects the presence of phonons as, according the Feynman relation $E(k) = \hbar^2k^2/[2mS(k)] = \hbar |k| c$, it corresponds to a linear excitation spectrum.
This eventually justifies the use of the Luttinger liquid theory which applies when the low-energy spectrum is gapless and linear.
For $^1$H, the low-momentum slope once expressed in natural units of $v_F$ decreases monotonously with the density, as can be appreciated from Fig.~\ref{Fig:Sk}a.
Together with an immediate formation of a peak at $|k|=2k_F$ this suggests that the hydrogen behaves similarly to a repulsive Fermi gas.
Instead, for deuterium and tritium (see Figs.~\ref{Fig:Sk}b-c) the slope first increases for densities up to $\rho\sim 0.03$~\AA$^{-1}$ and decreases later.
The initial increase in the slope is followed by the disappearance of the kink at $|k|=2k_F$ for $^{3}$H and a smooth featureless dependence on the momentum, typical for an interacting Bose gas.
Further increase in the density leads to formation of diverging peaks for all three isotopes, characteristic of the quasi-crystal regime.
The height of $\ell$-th peak diverges as $S(2\ell k_F)=A_{\ell} N^{1-2\ell^2 K}$ as can be obtained from Fourier transform of Eq.~(\ref{Eq:g2:LL}).
Instead, for a true crystal the height of the peak grows linearly with $N$.
Therefore, what we obtain in 1D is a weaker divergence with $N$ characteristic of a quasi-crystal.

\begin{figure}
\begin{center}
\begin{center}
\includegraphics[width=0.8\textwidth]{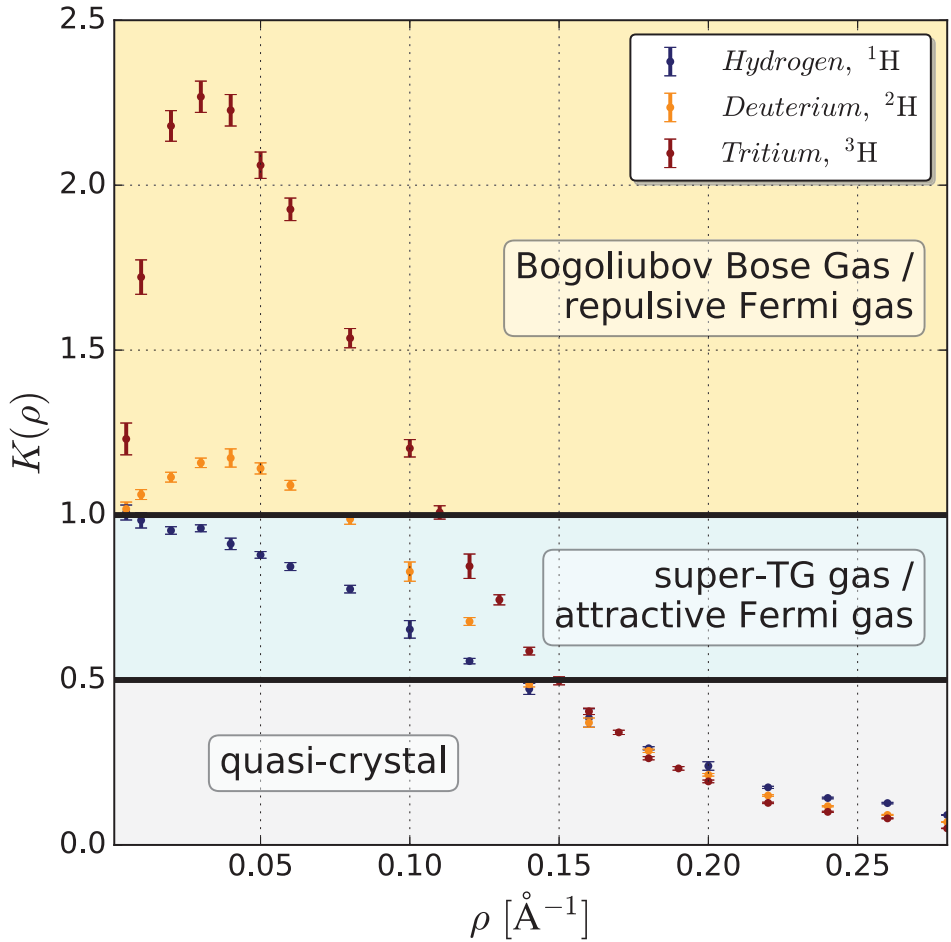}
\end{center}
\caption{Luttinger parameter $K$ as a function of the linear density for $^1$H, $^2$H, $^3$H
(increasing height of the peak) obtained from the linear slope~(\ref{Eq:K:Sk}) of $S(k)$ when $k \to 0$.
}
\label{Fig:K}
\end{center}
\end{figure}

We calculate the Luttinger parameter $K = v_F / c$ by extracting the speed of sound from the slope of the static structure factor
\begin{equation}
K = 2\pi\rho \lim_{k\to 0}\frac{S_k}{k}
\label{Eq:K:Sk}
\end{equation}
It can also be obtained from the fit to the two-body distribution function Eq.~(\ref{Eq:g2:LL}) and we verified that consistent values are obtained. Figure~\ref{Fig:K} shows the density dependence of the Luttinger parameter $K$.
Its knowledge permits us to distinguish different physical regimes.
While the system always remains in the non-superfluid non-condensed gas phase, still there are physically different regimes.
The $^1$H atoms, being the lightest isotope, manifest the strongest quantum effects originating from the largest quantum fluctuations.
As the density is increased, being a boson, it passes from the TG gas regime, $K=1$, to super-TG gas, $1/2<K<1$, and eventually to the quasi-crystal regime, $K<1/2$.
In this case, the density dependence of $K$ is monotonous.
The $^2$H isotope is a fermion and it passes from ideal fermions, $K=1$, to the attractive Fermi gas regime, $K>1$.
At higher densities there is a non-monotonous dependence of $K$, so that at some critical density ($\rho \approx 0.07$~\AA$^{-1}$) the gas again has the same Luttinger parameter as in an ideal Fermi gas.
This means that as far as the long-range response is concerned, it is similar to that of ideal fermions.
Also the renormalization group (RG) conclusions concerning the behavior of the system in presence of an external field of a certain type remain the same as for an IFG.
On the other hand, while the oscillations in the two-body distribution function $g(x)$ decay with the same power-law, the
amplitudes $A_{\ell}$ of oscillations in Eq.~(\ref{Eq:g2:LL}) are different.
At even larger densities, the correlations are stronger than in the IFG similarly to a repulsive Fermi gas.
For very large densities deuterium enters the quasi-crystal regime, in which we can note that the difference between different isotopes becomes relatively small.
The $^3$H isotope is the heaviest one and it possesses, probably, the most interesting phase diagram.
In this case the region of a TG gas is greatly reduced and is reached at densities much smaller compared to other
isotopes.
As the density is increased, tritium starts to behave similarly to a Bogoliubov gas of weakly interacting bosons.
The large maximal value of the Luttinger parameter in tritium, as compared to other isotopes, is a consequence of being close to the formation of a bound state, see Fig.~\ref{Fig:as(m)}. This is consistent with the equation of state exhibiting tendency to form an inflection,  at which point compressibility and the Luttinger parameter go to infinity.
In the point where the bound state enters, its energy is equal to zero, which is reminiscent of an ideal Bose gas with $K\to\infty$. 
Interestingly, the maximal value of the Luttinger parameter, reached at $\rho\approx 0.3$~\AA$^{-1}$, is above both the critical value of the localization in random disorder, $K=3/2$~\cite{Giamarchi88,Ristivojevic12} and the critical value for the pinning transition in commensurate optical lattices, $K=2$~\cite{Bloch08}.
Together with super-TG and quasi-crystal regimes at high densities, this makes the physical description of the accessible regimes very rich.

\section{Discussion\label{Sec:Discussion}}

It is instructive to compare the properties of hydrogen with those of other gases confined to a one-dimensional geometry.
The interactions between dilute alkali gases can be well approximated by a delta pseudopotential resulting
in the Lieb-Liniger model\cite{LiebLiniger63}, which features a crossover from the weakly-interacting Gross-Pitaevskii regime ($K\to\infty$) to the Tonks-Girardeau gas ($K=1$).
This crossover was experimentally observed\cite{Paredes04,Kinoshita04,Tolra04,Haller11} by tuning the interaction strength using the Olshanii confinement induced resonance (CIR).
The super-Tonks-Girardeau regime with $K<1$ corresponds to a metastable state which was experimentally realized by a fast sweep across the CIR\cite{Haller09}.
The energetic properties of mesoscopic two-component Fermi gases with a tunable $s$-wave interaction were measured using RF spectroscopy\cite{Serwane11,Wenz13}.
Bosons, interacting with a non-integrable repulsive interaction at short distances have the same energetic properties as fermions with the same interaction, according to Girardeau's mapping.
As a result, dipoles\cite{Arkhipov05, Citro07, Astrakharchik08} and hard rods\cite{Mazzanti08, Mazzanti08b} form a
Tonks-Girardeau/ideal Fermi gas at small density, pass through super-Tonks-Girardeau phase and form a quasi-crystal at large densities.
On the opposite, for Coulomb charges, the Wigner quasi-crystal is formed at low densities and TG/IFG at large ones\cite{Astrakharchik11,Ferre15}.
Calogero-Sutherland model permits to access all regimes with $K>0$\cite{Astrakharchik06}.

As concerning a comparison with other light elements, helium also has an interaction potential of van der Waals type.
For $^{4}$He data are available only for selected densities in narrow quasi-1D nanopores, correspoding to quasi-crystal regime~\cite{DelMaestro11,Kulchytskyy,VranjesGlyde}.
The density dependence of the Luttinger parameter in $^3$He in 1D \cite{Astrakharchik14} is overall quite similar to that of deuterium and tritium, with a similar location of the maximum, at $\rho \approx 0.05$\AA$^{-1}$.
We find that the hydrogen, being the lightest atom in the periodic table, shows a dramatic dependence on the isotope mass.

\section{Conclusions\label{Sec:Conclusions}}

To conclude, we studied how the ground-state properties of one-dimensional hydrogen are affected by the isotope mass.
The limit of ultra-low density corresponds to an ideal Fermi gas for Fermi-Dirac statistics and to a Tonks-Girardeau gas for Bose-Einstein case.
At high density, a quasi-crystal is formed as manifested by strong oscillations in the two-body distribution function $g(r)$ and diverging peak in the static structure factor $S(k)$.
We extract the Luttinger parameter $K$ from the linear behavior of $S(k)$ when $k \to 0$.
Based on a specific value of $K$ we define different physical regimes including Tonks-Girardeau, Bogoliubov Bose, super-Tonks-Girardeau gases and quasi-crystal for bosons; and ideal Fermi gas, attractive and repulsive Fermi gas, quasi-crystal regimes for fermions.
A peculiarity of one-dimensional hydrogen is that due to Girardeau's mapping for hard-core interactions, the energy and diagonal properties depend rather on mass and not on the Bose-Einstein or Fermi-Dirac statistics.
The isotope mass plays a non-straightforward role making the three species intrinsically different: the resulting $s$-wave scattering length $a_s$ is positive for $^1$H leading to positive corrections to Tonks-Girardeau energy (hard-rods or super-Tonks-Girardeau type); $a_s$ is negative for $^2$H and $^3$H causing negative corrections (Lieb-Liniger type). 
The most significant differences are observed in the case of tritium, where the mass is close to the value needed to form a bound state, causing large maximal values of the Luttinger parameter.

\section*{Acknowledgements}
We acknowledge partial financial support from the MICINN (Spain) Grant No.~FIS2014-56257-C2-1-P. 
L.V.M. acknowledges partial financial support by the Croatian Science Foundation under the project number IP-2014-09-2452. 
The Barcelona Supercomputing Center (The Spanish National Supercomputing Center -- Centro Nacional de Supercomputaci\'on) is
acknowledged for the provided computational facilities.

\section*{Appendix: Harmonic crystal theory}

A quasi-crystal is formed in the  high-density regime.
Its energy can be compared to that of a perfect crystal with particle position equally separated, $x_n = n/\rho$:
\begin{equation}
\frac{E^{(0)}}{N} = \sum\limits_{n=1}^\infty V(n/\rho).
\label{Eq:HC0}
\end{equation}

The excitation spectrum $\omega(k)$ can be obtained within harmonic crystal theory as a summation of the Hessian matrix over the perfect lattice
\begin{equation}
\omega(k)=\sqrt{\frac{4}{m}\sum_{n=1}^{\infty} \left[\frac{\partial^2V}{\partial x^2}\bigg|_{\frac{n}{\rho}} \sin^2\left(\frac{k n}{2\rho}\right)\right]}
\label{Eq:omega}
\end{equation}
The correction to the energy~(\ref{Eq:HC1}) can be obtained by integrating the phonon energy $\hbar\omega(k)/2$ over the first Brillouin zone (BZ)
\begin{equation}
\frac{E^{(1)}}{N}=\frac{1}{\ell_{BZ}}\int_{BZ}\frac{\hbar\omega(k)}{2}\;dk \; ,
\label{Eq:HC1}
\end{equation}
For the JDW potential, the summation~(\ref{Eq:HC0}-\ref{Eq:omega}) and integration~(\ref{Eq:HC1}) can be performed numerically.
Alternatively, for extremely large densities where the repulsive part of the interaction potential becomes dominant, some useful expressions can be obtained analytically.
The strong repulsion at short distances ($|x| \lesssim 2$\AA) can be approximated by an exponential wall
\begin{equation}
V(x) = V_0 \exp(-\varkappa |x|),
\label{Eq:Vint:exponential}
\end{equation}
with $V_0 = 4.73\times 10^5$ K and $\varkappa = 2.52 $\AA$^{-1}$.
The ground-state energy of a classic crystal~(\ref{Eq:HC0}) is the sum of the potential energy over a perfect lattice
\begin{equation}
\frac{E^{(0)}}{N}
= \frac{1}{2}\sum\limits_{n\ne 0} V(n/\rho)
= \frac{V_0}{e^{\varkappa/\rho}-1}
\end{equation}
as we report in Eq.~(\ref{Eq:HC0:exp})

The excitation spectrum is
\begin{equation}
\omega(k) =
\sqrt{
\frac{V_0\varkappa^2}{m} \frac{e^{\varkappa/\rho}+1}{e^{\varkappa/\rho}-1}
\frac{1-\cos(k/\rho)}{\cosh(\varkappa/\rho) -\cos(k/\rho)}
}
\approx
\sqrt{
\frac{4V_0\varkappa^2}{m e^{\varkappa/\rho}}}
\left|\sin\frac{k}{2\rho}\right|
\label{Eq:HC:omega}
\end{equation}
The spectrum is linear at small momenta, $\omega(k) = c|k|$, is defined on the first Brillouin zone, $-\pi\rho<k<\pi\rho$, and is a periodic function.

Finally, the energy correction~(\ref{Eq:HC1}) due to phonons is
\begin{equation}
\frac{E^{(1)}}{N}
=
\sqrt{\frac{V_0\hbar^2\varkappa^2}{m} \frac{e^{\varkappa/\rho}+1}{e^{\varkappa/\rho}-1}}
\left(
1
-
\frac{2}{\pi} \arctan(e^\frac{\varkappa}{2\rho})
\right)
\approx
\sqrt{\frac{4V_0\hbar^2\varkappa^2}{\pi^2m}}
e^{-\frac{1}{2}\frac{\varkappa}{\rho}}
\label{Eq:HC1:exp}
\end{equation}

\section*{Bibliography}
\bibliographystyle{ieeetr}


\begin{thebibliography}{10}

\bibitem{BEC95Cornell}
M.~H. Anderson, J.~R. Ensher, M.~R. Matthews, C.~E. Wieman, and E.~A. Cornell,
  ``Observation of Bose-Einstein condensation in a dilute atomic vapor,'' {\em
  Science}, vol.~269, no.~5221, pp.~198--201, 1995.

\bibitem{BEC95Ketterle}
K.~B. Davis, M.~O. Mewes, M.~R. Andrews, N.~J. van Druten, D.~S. Durfee, D.~M.
  Kurn, and W.~Ketterle, ``Bose-Einstein condensation in a gas of sodium
  atoms,'' {\em Phys. Rev. Lett.}, vol.~75, pp.~3969--3973, Nov 1995.

\bibitem{Girardeau60}
M.~Girardeau, ``Relationship between systems of impenetrable bosons and
  fermions in one dimension,'' {\em Journal of Mathematical Physics}, vol.~1,
  no.~6, pp.~516--523, 1960.

\bibitem{Astrakharchik05}
G.~E. Astrakharchik, J.~Boronat, J.~Casulleras, and S.~Giorgini, ``Beyond the
  tonks-girardeau gas: Strongly correlated regime in quasi-one-dimensional bose
  gases,'' {\em Phys. Rev. Lett.}, vol.~95, p.~190407, Nov 2005.

\bibitem{Olshanii98}
M.~Olshanii, ``Atomic scattering in the presence of an external confinement and
  a gas of impenetrable bosons,'' {\em Phys. Rev. Lett.}, vol.~81,
  pp.~938--941, Aug 1998.

\bibitem{Kinoshita04}
T.~Kinoshita, T.~Wenger, and D.~S. Weiss, ``Observation of a one-dimensional
  tonks-girardeau gas,'' {\em Science}, vol.~305, no.~5687, pp.~1125--1128,
  2004.

\bibitem{Paredes04}
A.~Paredes, Belen~Widera, V.~Murg, O.~Mandel, S.~Folling, I.~Cirac, G.~V.
  Shlyapnikov, T.~W. Hansch, and I.~Bloch, ``Tonks-girardeau gas of ultracold
  atoms in an optical lattice,'' {\em Nature}, vol.~429, pp.~277--281, Apr
  2004.

\bibitem{Haller09}
E.~Haller, M.~Gustavsson, M.~J. Mark, J.~G. Danzl, R.~Hart, G.~Pupillo, and
  H.-C. N\"gerl, ``Realization of an excited, strongly correlated quantum gas
  phase,'' {\em Science}, vol.~325, no.~5945, pp.~1224--1227, 2009.

\bibitem{Haller11}
E.~Haller, M.~Rabie, M.~J. Mark, J.~G. Danzl, R.~Hart, K.~Lauber, G.~Pupillo,
  and H.-C. N\"agerl, ``Three-body correlation functions and recombination
  rates for bosons in three dimensions and one dimension,'' {\em Phys. Rev.
  Lett.}, vol.~107, p.~230404, Dec 2011.

\bibitem{Kleppner98}
D.~G. Fried, T.~C. Killian, L.~Willmann, D.~Landhuis, S.~C. Moss, D.~Kleppner,
  and T.~J. Greytak, ``Bose-Einstein condensation of atomic hydrogen,'' {\em
  Phys. Rev. Lett.}, vol.~81, pp.~3811--3814, Nov 1998.

\bibitem{blume} D. Blume, B. D. Esry, Chris H. Greene, N. N. Klausen,
and G. J. Hanna, ``Formation of Atomic Tritium Clusters and Bose-Einstein Condensates,'' {\em Phys. Rev. Lett.}, vol.~89, pp.~163402, 2002.

\bibitem{kolos1968}
W.~Kolos and L.~Wolniewicz, ``Improved theoretical ground-state energy of the
  hydrogen molecule,'' {\em The Journal of Chemical Physics}, vol.~49, no.~1,
  pp.~404--410, 1968.

\bibitem{POT2}
M.~Jamieson, A.~Dalgarno, and L.~Wolniewicz, ``Calculation of properties of
  two-center systems,'' {\em Physical Review A}, vol.~61, no.~4, p.~042705,
  2000.

\bibitem{yan}Zong-Chao Yan, James F. Babb, A. Dalgarno, and G. W. F. Drake, ``Variational calculations of dispersion coefficients for interactions among H, He, and Li atoms,'' {\em Phys. Rev A},  vol.~54, p.~824, 1996.

\bibitem{potboro}
L.~V. Marki{\'c}, J.~Boronat, and J.~Casulleras, ``Quantum monte carlo
  simulation of spin-polarized H,'' {\em Physical Review B}, vol.~75, no.~6,
  p.~064506, 2007.

\bibitem{HDT3}
I.~Be{\v{s}}li{\'c}, L.~V. Marki{\'c}, and J.~Boronat, ``Spin-polarized
  hydrogen and its isotopes: A rich class of quantum phases (review article),''
  {\em Low Temperature Physics}, vol.~39, no.~10, pp.~857--873, 2013.

\bibitem{bevslic2009quantum}
I.~Be{\v{s}}li{\'c}, L.~V. Marki{\'c}, and J.~Boronat, ``Quantum Monte Carlo study of large spin-polarized tritium clusters,'' {\em The Journal of
  chemical physics}, vol.~131, no.~24, p.~244506, 2009.

\bibitem{xu2002structural}
H.~Xu and J.-P. Hansen, ``Structural and thermodynamic properties of
  spin-polarized fluid hydrogen,'' {\em Physics of Plasmas (1994-present)},
  vol.~9, no.~1, pp.~21--27, 2002.

\bibitem{bevslic2013quantum}
I.~Be{\v{s}}li{\'c}, L.~V. Marki{\'c}, J.~Casulleras, and J.~Boronat, ``Quantum
  Monte Carlo study of spin-polarized deuterium,'' {\em Physical Review B},
  vol.~88, no.~2, p.~024507, 2013.

\bibitem{Guardiola}
R.~Guardiola, {\em Monte Carlo techniques in the many body problem}.
\newblock Departament of Nuclear Physics, University of Granada.

\bibitem{boronat1994monte}
J.~Boronat and J.~Casulleras, ``Monte Carlo analysis of an interatomic
  potential for He,'' {\em Physical Review B}, vol.~49, no.~13, p.~8920, 1994.

\bibitem{Beslic08}
I.~Bešlić, L.~V.~Markić, and J.~Boronat, ``Quantum Monte Carlo study
  of small pure and mixed spin-polarized tritium clusters,'' {\em The Journal
  of Chemical Physics}, vol.~128, no.~6, 2008.

\bibitem{Beslic09}
I.~Bešlić, L.~V. Markić, and J.~Boronat, ``Stability limits of mixed
  spin-polarised tritium clusters,'' {\em Journal of Physics: Conference
  Series}, vol.~150, no.~3, p.~032010, 2009.

\bibitem{Stipanovic11}
P.~Stipanović, L.~V. Markić, J.~Boronat, and B.~Kežić, ``Ground state of
  small mixed helium and spin-polarized tritium clusters: A quantum Monte Carlo
  study,'' {\em The Journal of Chemical Physics}, vol.~134, no.~5, 2011.

\bibitem{Beslic13}
I.~Bešlić, L.~V. Marki\'c, and J.~Boronat, ``Spin-polarized hydrogen and
  its isotopes: A rich class of quantum phases (review article),'' {\em Low
  Temperature Physics}, vol.~39, no.~10, pp.~857--873, 2013.

\bibitem{Reatto67}
L.~Reatto and G.~V. Chester, ``Phonons and the properties of a bose system,''
  {\em Phys. Rev.}, vol.~155, pp.~88--100, Mar 1967.

\bibitem{Haldane81}
F.~D.~M. Haldane, ``Effective harmonic-fluid approach to low-energy properties
  of one-dimensional quantum fluids,'' {\em Phys. Rev. Lett.}, vol.~47,
  pp.~1840--1843, Dec 1981.

\bibitem{pures}
J.~Casulleras and J.~Boronat, ``Unbiased estimators in quantum Monte Carlo
  methods: Application to liquid $^{4}\mathrm{He}$,'' {\em Phys. Rev. B},
  vol.~52, pp.~3654--3661, Aug 1995.

\bibitem{tritiumleandra}
I.~Be\ifmmode \check{s}\else \v{s}\fi{}li\ifmmode~\acute{c}\else \'{c}\fi{},
  L.~V. Marki\ifmmode~\acute{c}\else \'{c}\fi{}, and J.~Boronat,
  ``Quantum Monte Carlo simulation of spin-polarized tritium,'' {\em Phys. Rev.
  B}, vol.~80, p.~134506, Oct 2009.
  
\bibitem{PhysRevA.81.013612}
G.E.~Astrakharchik, J.~Boronat, I.L.~Kurbakov, Yu.E. Lozovik,~F.Mazzanti,
``Low-dimensional weakly interacting Bose gases: Nonuniversal equations of state,''
{\em Phys. Rev. A}, vol.~81, p.~013612, Jan 2010.  

\bibitem{Arkhipov05}
A.~Arkhipov, G.~Astrakharchik, A.~Belikov, and Y.~Lozovik, ``Ground-state
  properties of a one-dimensional system of dipoles,'' {\em Journal of
  Experimental and Theoretical Physics Letters}, vol.~82, no.~1, pp.~39--43,
  2005.

\bibitem{Astrakharchik08}
G.~E. Astrakharchik and Y.~E. Lozovik, ``Super-Tonks-Girardeau regime in
  trapped one-dimensional dipolar gases,'' {\em Phys. Rev. A}, vol.~77,
  p.~013404, Jan 2008.

\bibitem{Astrakharchik09}
G.~E. Astrakharchik, G.~D. Chiara, G.~Morigi, and J.~Boronat, ``Thermal and
  quantum fluctuations in chains of ultracold polar molecules,'' {\em Journal
  of Physics B: Atomic, Molecular and Optical Physics}, vol.~42, no.~15,
  p.~154026, 2009.

\bibitem{Astrakharchik06}
G.~E. Astrakharchik, D.~M. Gangardt, Y.~E. Lozovik, and I.~A. Sorokin,
  ``Off-diagonal correlations of the Calogero-Sutherland model,'' {\em Phys.
  Rev. E}, vol.~74, p.~021105, Aug 2006.

\bibitem{Astrakharchik11}
G.~E. Astrakharchik and M.~D. Girardeau, ``Exact ground-state properties of a
  one-dimensional Coulomb gas,'' {\em Phys. Rev. B}, vol.~83, p.~153303, Apr
  2011.

\bibitem{Astrakharchik14}
G.~E. Astrakharchik and J.~Boronat, ``Luttinger-liquid behavior of
  one-dimensional $^{3}\mathrm{He}$,'' {\em Phys. Rev. B}, vol.~90, p.~235439,
  Dec 2014.

\bibitem{Glazman92}
L.~I. Glazman, I.~M. Ruzin, and B.~I. Shklovskii, ``Quantum transport and
  pinning of a one-dimensional Wigner crystal,'' {\em Phys. Rev. B}, vol.~45,
  pp.~8454--8463, Apr 1992.

\bibitem{Kane92}
C.~L. Kane and M.~P.~A. Fisher, ``Transport in a one-channel Luttinger
  liquid,'' {\em Phys. Rev. Lett.}, vol.~68, pp.~1220--1223, Feb 1992.

\bibitem{Furusaki93}
A.~Furusaki and N.~Nagaosa, ``Single-barrier problem and Anderson localization
  in a one-dimensional interacting electron system,'' {\em Phys. Rev. B},
  vol.~47, pp.~4631--4643, Feb 1993.

\bibitem{Giamarchi88}
T.~Giamarchi and H.~J. Schulz, ``Anderson localization and interactions in
  one-dimensional metals,'' {\em Phys. Rev. B}, vol.~37, pp.~325--340, Jan
  1988.

\bibitem{Ristivojevic12}
Z.~Ristivojevic, A.~Petkovi\ifmmode~\acute{c}\else \'{c}\fi{}, P.~Le~Doussal,
  and T.~Giamarchi, ``Phase transition of interacting disordered bosons in one
  dimension,'' {\em Phys. Rev. Lett.}, vol.~109, p.~026402, Jul 2012.

\bibitem{Bloch08}
I.~Bloch, J.~Dalibard, and W.~Zwerger, ``Many-body physics with ultracold
  gases,'' {\em Rev. Mod. Phys.}, vol.~80, pp.~885--964, Jul 2008.

\bibitem{LiebLiniger63}
E.~H. Lieb and W.~Liniger, ``Exact analysis of an interacting bose gas. i. the
  general solution and the ground state,'' {\em Phys. Rev.}, vol.~130,
  pp.~1605--1616, May 1963.

\bibitem{Tolra04}
B.~L. Tolra, K.~M. O'Hara, J.~H. Huckans, W.~D. Phillips, S.~L. Rolston, and
  J.~V. Porto, ``Observation of reduced three-body recombination in a
  correlated 1d degenerate bose gas,'' {\em Phys. Rev. Lett.}, vol.~92,
  p.~190401, May 2004.

\bibitem{Serwane11}
F.~Serwane, G.~Z\"urn, T.~Lompe, T.~B. Ottenstein, A.~N. Wenz, and S.~Jochim,
  ``Deterministic preparation of a tunable few-fermion system,'' {\em Science},
  vol.~332, no.~6027, pp.~336--338, 2011.

\bibitem{Wenz13}
A.~N. Wenz, G.~Z\"urn, S.~Murmann, I.~Brouzos, T.~Lompe, and S.~Jochim, ``From
  few to many: Observing the formation of a fermi sea one atom at a time,''
  {\em Science}, vol.~342, no.~6157, pp.~457--460, 2013.

\bibitem{Citro07}
R.~Citro, E.~Orignac, S.~De~Palo, and M.~L. Chiofalo, ``Evidence of
  Luttinger-liquid behavior in one-dimensional dipolar quantum gases,'' {\em
  Phys. Rev. A}, vol.~75, p.~051602, May 2007.

\bibitem{Mazzanti08}
F.~Mazzanti, G.~E. Astrakharchik, J.~Boronat, and J.~Casulleras, ``Ground-state
  properties of a one-dimensional system of hard rods,'' {\em Phys. Rev.
  Lett.}, vol.~100, p.~020401, Jan 2008.

\bibitem{Mazzanti08b}
F.~Mazzanti, G.~E. Astrakharchik, J.~Boronat, and J.~Casulleras, ``Off-diagonal
  ground-state properties of a one-dimensional gas of Fermi hard rods,'' {\em
  Phys. Rev. A}, vol.~77, p.~043632, Apr 2008.

\bibitem{Ferre15}
G.~Ferr\'e, G.~E. Astrakharchik, and J.~Boronat, ``Phase diagram of a quantum
  coulomb wire,'' {\em Phys. Rev. B}, vol.~92, p.~245305, Dec 2015.

\bibitem{DelMaestro11}
A.~Del~Maestro, M.~Boninsegni, and I.~Affleck, ``$^{4}\mathrm{He}$ Luttinger
  liquid in nanopores,'' {\em Phys. Rev. Lett.}, vol.~106, p.~105303, Mar 2011.

\bibitem{Kulchytskyy}
B.~Kulchytskyy, G.~Gervais, and A.~Del~Maestro, ``Local superfluidity at the
  nanoscale,'' {\em Phys. Rev. B}, vol.~88, p.~064512, Aug 2013.

\bibitem{VranjesGlyde}
L.~V. Marki\ifmmode~\acute{c}\else \'{c}\fi{} and H.~R. Glyde,
  ``Superfluidity, BEC, and dimensions of liquid $^{4}\mathrm{He}$ in
  nanopores,'' {\em Phys. Rev. B}, vol.~92, p.~064510, Aug 2015.

\end{thebibliography}

\end{document}